\documentclass[fleqn,usenatbib]{mnras}
\usepackage{amsmath,amssymb}
\usepackage{mathptmx}
\usepackage{txfonts}

\usepackage[T1]{fontenc}

\DeclareRobustCommand{\VAN}[3]{#2}
\let\VANthebibliography\thebibliography
\def\thebibliography{\DeclareRobustCommand{\VAN}[3]{##3}\VANthebibliography}

\usepackage{graphicx}	

 \usepackage{chemformula}

\newcommand{\ind}[1]{_{\rm #1}}

\newcommand{\cc}[0]{cm$^{-3}$}
\newcommand{\kms}[0]{kms$^{-1}$}

   \title{Implications from secondary emission from neutral impact on Cassini plasma and dust measurements}
  \author[F. L. Johansson]{
  F. L. Johansson$^{1}$ \thanks{E-mail: \href{mailto:frejon@gmail.com}{frejon@gmail.com} FLJ}%
E. Vigren,$^{1}$
J.H. Waite,$^{2}$
K. Miller,$^{2}$
A. I. Eriksson,$^{1}$
N. J. T. Edberg,$^{1}$
J. Dreyer.$^{1}$
\\
$^{1}$Swedish Institute of Space Physics, Uppsala, Sweden, \\
$^{2}$Southwest Research Institute, San Antonio, TX, USA\\
}

\date{Accepted XXX. Received YYY; in original form ZZZ}

\pubyear{2021}

\begin{document}
\label{firstpage}
\pagerange{\pageref{firstpage}--\pageref{lastpage}}
\maketitle

%
%

\begin{abstract}
We investigate the role of secondary electron and ion emission from impact of gas molecules on the Cassini Langmuir Probe (RPWS-LP, or LP) measurements in the ionosphere of Saturn. We add a model of the emission currents, based on laboratory measurements and data from comet 1P/Halley, to the equations used to derive plasma parameters from LP bias voltage sweeps. Reanalysing several hundred sweeps from the Cassini Grand Finale orbits, we find reasonable explanations for three open conundrums from previous LP studies of the Saturn ionosphere. We find an explanation for the observed positive charging of the Cassini spacecraft, the possibly overestimated ionospheric electron temperatures, and the excess ion current reported. For the sweeps analysed in detail, we do not find (indirect or direct) evidence of dust having a significant charge-carrying role in Saturn’s ionosphere. We also produce an estimate of \ch{H_2O} number density from the last six revolutions of Cassini through Saturn’s ionosphere in higher detail than reported by the Ion and Neutral Mass Spectrometer (INMS). Our analysis reveals an ionosphere that is highly structured in latitude across all six final revolutions, with mixing ratios varying with two orders of magnitude in latitude and one order of magnitude between revolutions and altitude. The result is generally consistent with an empirical photochemistry model balancing the production of \ch{H+} ions with the \ch{H+} loss through charge transfer with e.g., \ch{H_2O}, \ch{CH_4} and \ch{CO_2}, for which water vapour appears as the likeliest dominant source of the signal in terms of yield and concentration.

 \end{abstract}



   \maketitle

\section{Introduction}
During its final orbits around Saturn, the Cassini spacecraft provided the first in-situ measurements of the ionosphere of the giant planet. The speed at periapsis must necessarily be high for a spacecraft orbiting a massive object like Saturn in an elliptic orbit, and Cassini plunged through the ionosphere at more than 30~km/s. This high flyby speed has implications for some payload instruments. For example, the high kinetic energy of atmospheric molecules as seen from the moving spacecraft limited the highest ion mass accessible to the Ion and Neutral Mass Spectrometer (INMS) to 8~Da and heavy neutrals and grains likely fragmented inside the instrument antechamber or adsorbed to the walls \citep{Cravens_ion_comp_2019,miller_cassini2020}.
In this paper, we present a study of another effect of the high flyby speed and its consequences for the measurements by the Langmuir probe of the Radio and Plasma Waves Science investigation (RPWS-LP). At 30~km/s, gas particles impact on the spacecraft surfaces at an energy of about 5~eV/Da. For sufficiently heavy atoms and molecules, the energy will exceed the typical ionisation energy ($\sim$ 10~eV) of the impacting particle as well as the work function of the surface material. One or more electrons, known as secondary electrons, may then be released into space. Ions can also be emitted, though typically at much lower flux \citep{schmidtarends1985See}. This emission contributes to the current measured by the LP and can thus be detected. We will attempt to identify this current in the LP data and to use it as an independent estimate of the content of heavy molecules, with very different limitations from the INMS measurements limitations as noted above. 

Because of the much lower speed ($\sim7$~km/s) of a satellite in low Earth orbit, secondary emission by molecular impact is negligible for spacecraft in the terrestrial ionosphere. Data from Earth satellites can therefore not be used for comparison to Cassini measurements. However, the flybys of comet 1P/Halley in 1986 occurred at very high relative speed, just below 70~km/s, providing the water molecules abundant in the cometary coma with an impact energy of about 500~eV, clearly sufficient for secondary emission. \citet{grard_observations_1989} found that the induced secondary emission from neutral impact was on such a scale that it dominated all other currents to the Giotto and Vega spacecraft. In contrast to the comet case, water is only a minor species in Saturn's upper atmosphere, and this is true also for other sufficiently heavy molecules like \ch{CO_2}. In addition, the relevant speed for Cassini is less than half of the comet case. Nevertheless, such molecules are still present and their impact energy of $\sim100$~eV is well sufficient for secondary emission, as shown also by the laboratory measurements acquired in support of the data analysis from Vega and Giotto \citep{schmidtarends1985See}.

The RPWS-LP data from the Cassini Grand Finale orbits have been analysed and presented in previous work by e.g.\ \citet{Wahlund2018science}, \citet{hadid2019saturn} and \citet{Morooka2019_saturnionosphere}. In particular, \citet{Morooka2019_saturnionosphere} (hereafter: M2019) suggested that a detected large negative  current to the LP was due to a population of positive ions with density much above that of the electrons. The situation showed obvious similarities to LP observations in the Enceladus plume, where negatively charged dust grains in the sub-micrometer size range (referred to as "nanograins") were invoked to explain the apparent charge imbalance \citep{wahlund_detection_2009,morooka_dusty_2011,shafiq_characteristics_2011}. Dust is obviously an important feature of the Enceladus plume, which is a complex mix of electrons, cations, anions and charged dust (both positive and negative), and this interpretation of the LP data is consistent with independent observations by other Cassini instruments \citep{jones_fine_2009,Coates2010negativeion,farrell_modification_2010,hill_charged_2012,engelhardt_plasma_2015,Postberg2018Nature_enc}. Suggesting a similar interpretation also for the apparent ion-electron charge imbalance in Saturn ionosphere was therefore an obvious interpretation, put forward by M2019. However, as the secondary current will give a contribution of the same kind as an ion current and also may be suspected to be more important in the Saturn ionosphere than at Enceladus (much lower flyby speed, typically 7-15~km/s, and in most cases lower neutral gas density) an investigation of its possible effects is merited. 
This is the topic of the present study, where we extend the analysis of M2019 by including a model for the current from secondary emission from neutral impact. We show that such model is often able to reproduce the main features of the LP sweeps in the Saturn ionosphere even without inclusion of a charged dust population, and also provide re-interpretations of some other details of the LP sweeps which are more inline with expectations from theory. While some features are better described by the dust interpretation of M2019, our results show that neglecting secondary emission from neutral impact likely overestimates the charged dust population in Saturn's ionosphere, sometimes very significantly. 

The paper is structured as follows. Section~\ref{sec:instr} briefly describes the instruments used. The standard model for interpretation of Langmuir probe bias sweeps is in Section~\ref{sec:method} extended to include secondary emission by neutral impact. We present the results of the analysis, and the results of the secondary emission model in terms of neutrals in Section~\ref{sec:results}. We discuss these results and go into details of the merits and drawbacks with our analysis results in comparison to M2019 in Section~\ref{sec:discussion}. Here we also discuss the implications of our measurements for the neutral composition at Saturn, before concluding in Section~ \ref{sec:conclusion}.

\section{Instruments}\label{sec:instr}

\subsection{RPWS}

The Cassini Radio and Plasma Wave Science investigation (\emph{RPWS}) was designed to study waves, thermal plasma, and dust in the vicinity of Saturn \citep{gurnett_cassini_2004}. We will make use of the data from the three 10~m long electric field antennas, with a tip-to-tip length of 18.52~m, which allows for the accurate determination of the plasma frequency (and thereby, the electron density) for a for a broad range of Debye lengths \citep{persoon_ne_2005}.

On a separate 0.8~m boom, RPWS also includes a 5~cm diameter Langmuir probe (\emph{LP}). The primary parameter measured is the current flowing to the probe when a bias voltage is applied to it. The current response from the plasma allows us to estimate ion and electron densities, solar EUV flux, spacecraft potential and electron temperatures.
Although the two techniques compliment each other regarding electron density determination, local disturbances such as spacecraft charging is much less likely to affect the determination of electron densities using electric field measurements \citep{Johansson_plasmaxcal_2021}, as RPWS samples a much larger volume than RPWS-LP. As such, the LP current at fixed bias voltage \citep[e.g.~][]{engelhardt_plasma_2015} or floating potential \citep[e.g.~][]{morooka_electron_2009} are generally converted (via a linear scaling) to electron densities by calibration to wave determined electron densities whenever available.

\subsection{INMS}

The Ion Neutral Mass Spectrometer (\emph{INMS}) is a quadrupole mass spectrometer capable of analyzing neutral compounds via both the open and closed sources, and ions via the open source \citep{waite_cassini_2004}. The relative spacecraft velocity with respect to Saturn during the proximal orbits exceeded 30\kms, and limited observable ion species to those with mass numbers less than 8~Da (i.e.\ only lighter ions). Also, at these speeds, heavier neutral species and grains break up in the closed source antechamber \citep{Teolis_encaladus_inms_2010}, leading to a possible overestimation of some of the fragmentation products and slight underestimation of larger molecules (organics).

\section{Method} \label{sec:method}

For the current measured by a Langmuir probe, an electron emitted by the probe is equivalent to a positive particle (ion) depositing its charge on the probe. Identification of electron emission is therefore important for estimates of ion density. For RPWS-LP there are two emission processes of interest. (1) photoelectrons emitted by electromagnetic radiation in the extreme ultraviolet range due to the photoelectric effect and (2) particles (such as molecules, ions, electrons) impacting at high velocities can transfer sufficient energy to allow one or more electrons to be excited and emitted, generally called secondary electrons. The work function of the material or the impacting molecule determines the minimum energy the process has to deliver for electron emission to occur. An added complexity for collisions of molecules on surfaces is that the conversion of impact energy to internal energy is often between 7 and 35 percent \citep{Laskin2003sid_energytransfer,meroueh_effect_2001}, with a typical reported value of 25 percent. 

The number of emitted particles per incoming particle, the quantum yield ($Y$), of these emission processes is material dependent and pertains mostly to the first 1~nm depth of the surface, making it sensitive to contamination/oxidation \citep{schmidtarends1985See,feuerbacher_experimental_1972,balcon_secondary_2010,pimpec_secondary_2003,samplon_secondary_2004}. For space-weathered metals, oxidation (during launch or otherwise) generally allows the quantum yield to not differ with much more than a factor of two between any two metals used in space applications \citep{grard_properties_1973}, and we will make use of the \citet{schmidtarends1985See} study of secondary electron emission of water molecules incident on space-weathered gold, but allow for errors of up to a factor two when extrapolating the results to the titanium nitride (TiN) coated Langmuir Probe. A similar approach was also used by \citet{johansson_rosetta_2017} to characterise the photoemission of an identical probe on the spacecraft Rosetta.

The current $I\ind{se0}$ generated by secondary electron emission from an impactor such as a water molecule, to an object of cross-sectional area $A$ is simply
\begin{equation}\label{eq:ise0}
    I\ind{se0} = v\ind{ram}n\ind{H_2O}\pi r\ind{LP}^2 eY\ind{H_2O}^e,
\end{equation}
where $v\ind{ram}$ is the impact velocity, $n\ind{H_2O}$ is the water gas number density, $r\ind{LP}$ is the radius of the probe, and $Y\ind{H_2O}^e$ is the quantum yield, the number of emitted electrons per incident water molecule at the kinetic energy of the impacting particle, and $e$ is the elementary charge. The process is of course not restricted to water molecules, but for the specific case of water gas impacting space-weathered metals, laboratory experiments report a quantum yield $Y\ind{H_2O}^e$ of 0.1-0.2 \citep{schmidtarends1985See}.

If the surface is charged, the current to or from the surface will depend on whether charged particles are attracted or repelled by it. For a Langmuir probe, where the potential is biased by a potential $V\ind{b}$ from the spacecraft ground, the potential of the probe with respect to surrounding space is
\begin{equation}\label{eq:vp}
V\ind{p} = V\ind{b}+V\ind{S},
\end{equation}
where $V\ind{S}$ is the voltage between the spacecraft and surrounding unperturbed space. For ionospheric plasma particles, it is this absolute potential of the probe that dictates if the particle is attracted or repelled by it. For locally produced electrons, such as photoelectrons or secondary electrons, it is the electric field of the plasma immediately surrounding the probe that determines the net force acting on the newly emitted electron. Therefore, at some bias potential where $V\ind{\dagger} = 0$ we shift from net repulsion to attraction. For a probe inside the electrostatic potential field of another body, if the absolute potential at the probe position is some factor $\alpha$ ($0 \leq \alpha \leq 1$)  of the spacecraft potential, we can define
\begin{equation} \label{eq:vdagger}
    V\ind{\dagger} = V\ind{b} + \alpha V\ind{S},
\end{equation}
in analogy to Eq~\ref{eq:vp}.

\citet{grard_properties_1973} describes the case of a Maxwell-Boltzmann distribution of photoelectrons emitted from a probe and shows that the emitted current from a plane source is

\begin{equation}\label{eq:iphisee}
    I\ind{ph} =
    	\begin{cases}
	        -I\ind{ph0}\exp\left(\frac{- e V\ind{\dagger}}{k\ind{B} T\ind{ph}}\right)
 & \text{for }  V\ind{\dagger} \geq 0 \\
    -I\ind{ph0}  & \text{for }  V\ind{\dagger} < 0,
	\end{cases}
\end{equation}
where $T\ind{ph}$ is the Maxwellian temperature of the emitted photoelectrons, $I\ind{ph0} = j\ind{ph}\pi r\ind{LP}^2$, where $j\ind{ph}$ is the photosaturation current density, which depends on the surface material as well as the solar radiation spectrum in extreme ultraviolet, and we have adapted the expression to accommodate our definition of $V\ind{\dagger}$ similar to \citet{johansson_rosetta_2017}. We use the usual sign convention of considering currents as positive when flowing from the probe to the plasma.

As the current-voltage relation for electrons emitted by the probe material should not strictly depend on whether it was excited by an impact from a neutral particle or a photon, we can similarly define the secondary electron current from neutral impact:
\begin{equation}\label{eq:ise}
    I\ind{se} =
    	\begin{cases}
	        -I\ind{se0}\exp\left(\frac{- e V\ind{\dagger}}{k\ind{B} T\ind{se}}\right)
 & \text{for }  V\ind{\dagger} \geq 0 \\
    -I\ind{se0}  & \text{for }  V\ind{\dagger} < 0,
	\end{cases}
\end{equation}
by substituting $I\ind{ph0}$ with $I\ind{se0}$ as defined by Eq~\ref{eq:ise0}, and $T\ind{ph}$ with the equivalent Maxwellian temperature of the emitted secondaries, $T\ind{se}$. 

That this analogy generally holds is apparent in lab measurements of emitted secondary electrons and ions from neutrals, shown by \citet{schmidtarends1985See} (Fig.~1). Furthermore, \citet{schmidtarends1985See} find that the yield of secondary ion emission from neutral impact is generally a factor 10 lower than the corresponding yield for electrons for the same material and neutral species. We therefore expect this current contribution to be small, but for Langmuir Probe sweep analysis, it may still be useful to consider. It is reasonable to consider that Eq.~\ref{eq:ise} will apply also to this current, after adjusting for the opposite sign of the positive ions according to
\begin{equation}\label{eq:isi}
    I\ind{si} =
    	\begin{cases}
	             I\ind{si0}  & \text{for }  V\ind{\dagger} \geq 0\\
 I\ind{si0}\exp\left(\frac{qV\ind{\dagger}}{k\ind{B} T\ind{si}}\right) & \text{for }  V\ind{\dagger} < 0,
	\end{cases}
\end{equation}
where $T\ind{si}$ is the Maxwellian temperature of the emitted ions, and we will allow $I\ind{si0}$ to be a factor 5-20 lower than $I\ind{se0}$ (see Eq. ~\ref{eq:ise0}), in accordance with \citet{schmidtarends1985See}.

The choice of a Maxwellian energy distribution in \citet{grard_properties_1973} relies on the assumption that the photoemission or secondary emission occurs at some non-negligible penetration depth in the surface. There, emitted electrons would collide sufficiently many times inside the material to achieve a random distribution when leaving the material (instead of the expected discrete energy levels from the quantum processes involved). Arguably, this could be less true for a larger particle such as a molecule, which would not be able to penetrate as deep into the material. 
The assumption of Maxwellian distributions of the emitted ions and electrons in Eq. ~\ref{eq:iphisee}-\ref{eq:isi} should therefore be seen as a sufficiently good approximation for parametrization of observed probe sweeps over a few volts rather than as an exact description over a wider energy range \citep{pedersen_solar_1995}.

These currents should be added to the currents collected by the LP from the ambient plasma. \citet{mott-smith_theory_1926} introduced a useful method for analysing the currents to a body immersed in plasma known as Orbital-Motion-Limited theory, OML, which assumes particle trajectories based solely on conservation of energy and angular momentum. This approach is applicable as long as the electric field from the probe does not decay too rapidly with distance, meaning the shielding effects of the plasma must not be too strong. This can therefore be adopted when the radius of the probe $r\ind{p}$ is sufficiently much smaller than the Debye length, $\lambda\ind{D}$, which is the characteristic length scale of the Debye shielding phenomenon, the innate ability of the plasma to screen potential differences. In the opposite case of Debye lengths being short compared to the probe size, Sheath Limited theory (SL) applies, as described by \citet{laframboise_theory_1966}, which we will consider in Appendix~\ref{sec:appendix}.


By assuming that the ionospheric electron population follows a Maxwell-Boltzmann energy distribution, the current to a probe can be directly calculated as the flux through that volume from the random thermal motion times the charge of the electrons. For a non-drifting electron population (where thermal motion is much larger than the drift velocity), the thermal current $I\ind{e0}$ to a probe at the same potential as the plasma is then
 \begin{equation}
     I\ind{e0} = 4 \pi r\ind{LP}^2 e n\ind{e} \sqrt{\frac{k\ind{B}T\ind{e}}{2\pi m\ind{e}}},
 \end{equation}
where $n\ind{e}$ is the electron density, $T\ind{e}$ is the Maxwellian electron temperature and other symbols have their usual meaning.

From OML, assuming all particles are non-magnetized and coming from a zero potential at infinity, it can be shown~\citep{mott-smith_theory_1926} that the electron current $I\ind{e}$ to a spherical probe at potential $V\ind{p}$ is
\begin{equation} \label{eq:ie}
    I_e =
\begin{cases}
I\ind{e0}\left(1 + \frac{ e V\ind{p}}{ k\ind{B}T\ind{e}}\right)  & \text{for } V\ind{p} \geq 0  \\
    I\ind{e0} \exp\left({\frac{e V\ind{p}}{k\ind{B}T\ind{e}}}\right) & \text{for } V_p < 0 .
	\end{cases}
\end{equation}

As ions  are heavier than electrons, their thermal speed is much lower than that of the electrons even if their temperatures are equal. The Cassini speed through the Saturn ionosphere was much higher than the thermal speed of the ions, so the ion flow in the spacecraft reference frame can safely be assumed to be supersonic. We can then simplify the ion current $I\ind{i0}$ to the probe when at the potential of the plasma by
\begin{equation} \label{eq:Ii0}
    I\ind{i0} = \pi r\ind{p}^2 q\ind{i} n\ind{i} u\ind{i},
\end{equation}
where $q\ind{i}$ is the ion charge, $n\ind{i}$ is the ion density and $u\ind{i}$ is the ram speed of the ions. It can be shown \citep{fahleson_ionospheric_1974} that the ion current $I\ind{i}$ to a probe at a general potential is then
\begin{equation}\label{eq:ion}
I\ind{i} =
	\begin{cases}
	- I\ind{i0}\left(1 - \frac{ e V\ind{p}}{ E\ind{i}}\right) & \text{for } V\ind{p} \leq E\ind{i}/e \\
    0 & \text{for } V\ind{p} > E\ind{i}/e,
	\end{cases}
\end{equation}
where $E\ind{i}$ is the kinetic energy $\frac{1}{2} m\ind{i}u\ind{i}^2$ of ions of mass $m\ind{i}$. For a plasma with multiple ion species, $m\ind{i}$ is instead the harmonic mean ion mass, also called the effective mass \citep{Holmberg2012_ion_torus,Shebanits2016titan_ion}. In the limit of large kinetic energy (compared to the sweep range of the probe, which is typically $\pm 4$~V in the deep ionosphere), the minimum density of ions or charged dust that can be detected above the nominal noise floor is about 10~cm$^{-3}$ in the Saturnian ionosphere, as the noise level of the instrument typically is $\lesssim0.1$~nA.

\section{Results}
\label{sec:results}

\begin{figure}
    \includegraphics[width=1.0\columnwidth]{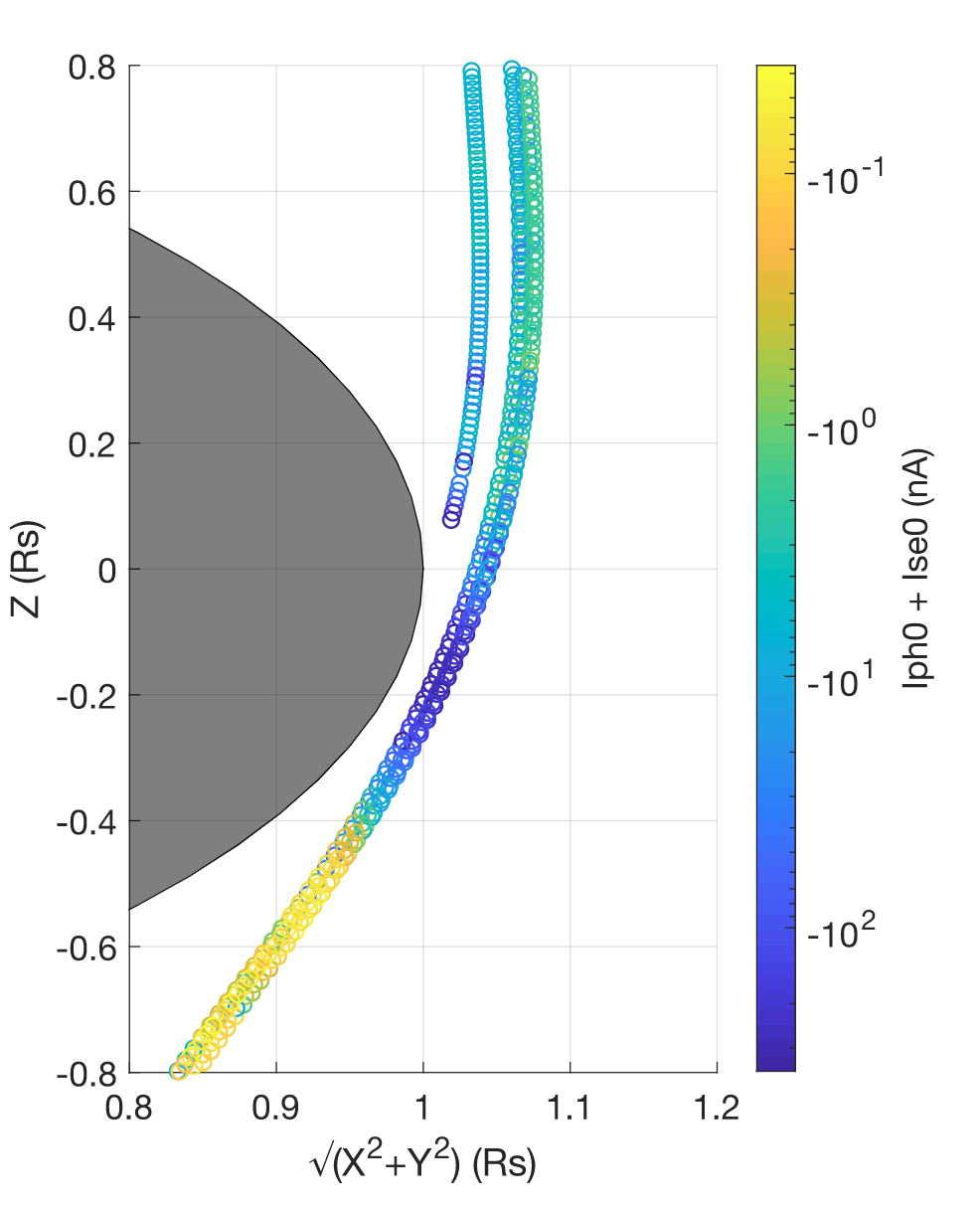}
    \caption{Cassini trajectories for the last six revolutions around Saturn, colour coded by the sum of the photoemission and secondary electron current as identified by the automatic routine. Saturn is drawn as a grey ellipsoid with an equatorial radius 1~$R\ind{s}$.}
    \label{fig:saturn_iph0}
\end{figure}

Photoemission and secondary electron emission from neutrals are effectively indistinguishable in a Langmuir Probe sweep analysis without prior assumptions on temperatures. Therefore, we applied a photoemission amplitude detection routine, validated in \citet{johansson_rosetta_2017} and described in the archive data documentation\footnote{\url{ftp://psa.esac.esa.int/pub/mirror/INTERNATIONAL-ROSETTA-MISSION/RPCLAP/RO-C-RPCLAP-5-PRL-DERIV2-V1.0/DOCUMENT/RO-IRFU-LAP-EAICD.PDF}} for an identical Langmuir probe aboard Rosetta, to analyse more than 19,500 RPWS-LP sweeps from the closest approaches of six of the last Cassini passes through the Saturn ionosphere (revolutions 288-293). Far from the Saturnian ionosphere, we recover the estimated photoemission saturation current of 0.5~nA, also reported in M2019. 
Further in, in the deepest part of the ionosphere as seen in Fig.\ref{fig:saturn_iph0}, we detect currents even exceeding 1~$\mu$A, which we will attempt to describe as secondary electron from neutral impact. We also find that during the outbound leg, both photoemission and secondary electron emission vanishes as expected in the shadows of the Saturnian rings in the absence of a neutral environment (at high altitudes), as studied by \citet{Hadid2018ring_shadowing}.

\begin{figure}
\centering
    \includegraphics[width=1.0\columnwidth]{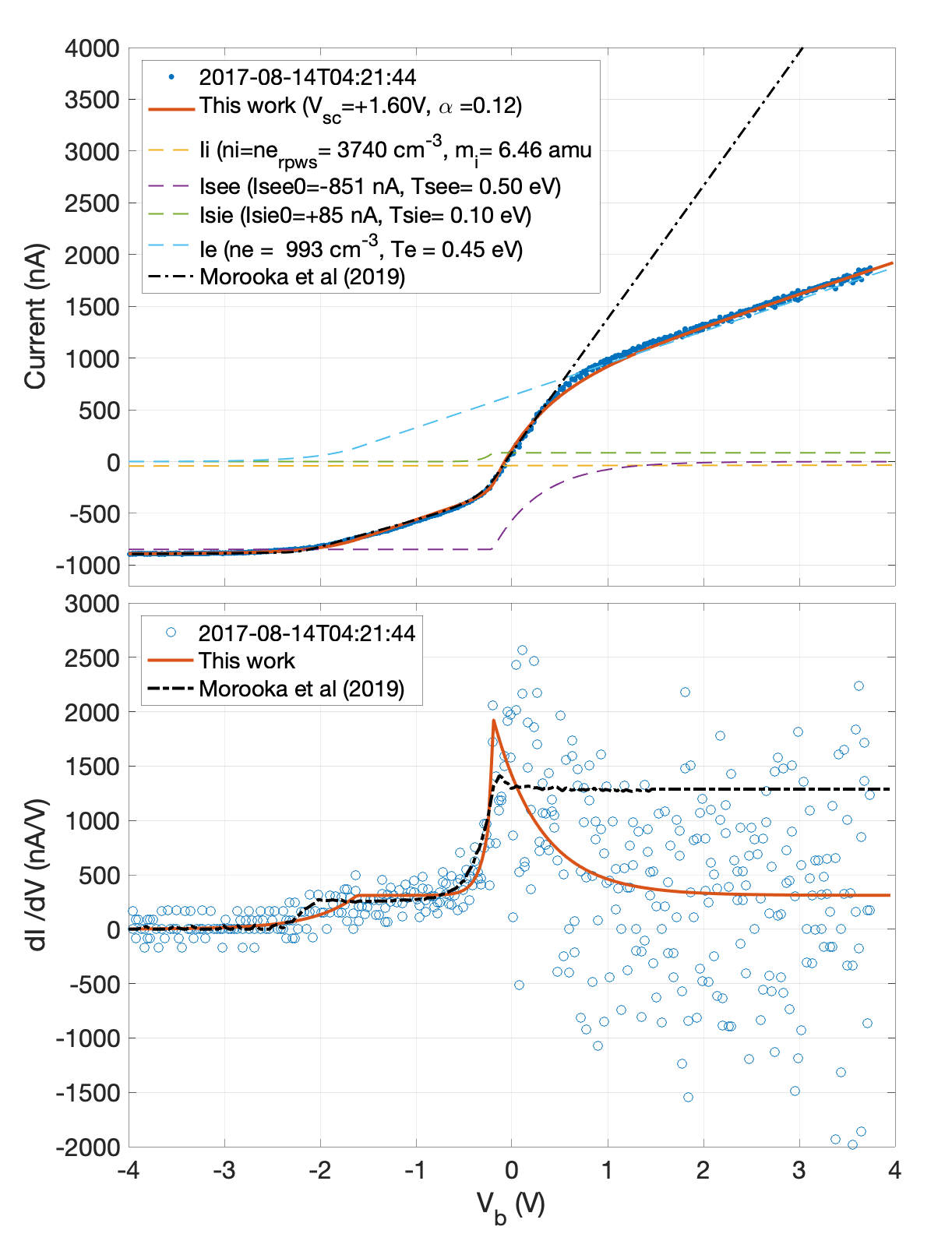}
    \caption{Currents and current derivatives vs bias potential, common x-axis.\textbf{Top:} Langmuir Probe sweep from 2017-08-14T04:21:44 from Rev. 288 (blue dots) and fitted sweep currents, including secondary electron and ion emission from neutrals (purple and green dashed line, respectively), electrons (blue dashed line), ions (yellow dashed line) and the total fit (red solid line) vs bias potential. Also plotted in black (dash-dotted line), for reference, the fit of the same sweep published in M2019. The harmonic mean of the ion mass from INMS (measuring ions up to 8~amu) at the time of the sweep was 1.49~amu. \textbf{Bottom:} The derivative of the sweep current (blue circles), and the derivative of the total fit of the analysis in this work (red line) and in M2019 (dashed black line) vs bias potential.}
    \label{fig:288_mich} 
\end{figure}

To further validate this routine, we provide detailed sweep analysis for a diverse set of sweeps. In Fig.~\ref{fig:288_mich}, we show a simple analysis using fits of four distinct currents/plasma populations. This particular sweep was also discussed in M2019 (Fig.~5). Apart from including secondary electron emission, we also attempt a fit of the whole sweep, while M2019 restrict to $V\ind{b} < 0.4$~V. By doing so, we find a fit that adheres much closer to the data at positive $V\ind{b}$. As we identify the majority of the detected (negative) current to be from secondary electron emission, the ion density estimate is a factor of 20 lower than the result in M2019 ($8.02\times10^{4}$~\cc).

This is just one example out of the 200 sweeps identified with significant secondary electron emission, including all sweeps below 2500~km altitude as seen in Fig.~\ref{fig:saturn_iph0}, all strongly correlating with neutral density. In Appendix~\ref{sec:appendix}, we present further examples, compare to the analysis in M2019 when possible, and test the validity of OML for modeling of RPWS-LP measurements by comparison to the numerical results by \citet{laframboise_theory_1966}. For all these 200 sweeps we find that the Debye length is always large enough that deviations from OML theory are insignificant. Also, the secondary electron/ion emission well explains the relaxation in the derivative, improving the residual of the fits by up two orders of magnitude and decreasing the ion density estimate significantly from to M2019.

\subsection{Neutrals}

There are several abundant gasses that we suspect can be responsible for secondary emission on the probe. The lightest and most abundant three gasses \ch{H,~H_2} and \ch{He}, can be ruled out by their low impact energy alone (owing to their small mass), and indeed the currents we see are several order of magnitudes lower than what otherwise would be expected. 
Our prime suspect responsible for secondary emission on the probe is instead water vapour. We know from laboratory measurements and cometary flybys that water vapour is efficient in producing secondary electron currents, is present in the Saturnian equatorial upper atmosphere, and that the impact energy is large enough to ionise the water molecule. We may expect other molecules to contribute to this signal in an additive behaviour, modulated by their respective yield. 

The low estimated mixing ratio for \ch{N_2} and the low quantum yield of \ch{CO_2} likely combine to reduce the signal from these molecules to approximately a third of \ch{H_2O} \citep{Moore2018modelsaturn,miller_cassini2020,schmidtarends1985See}. Methane may or may not trigger significant secondary electron emission at these energies, owing to its lower impact energy, stability and unknown quantum yield. Lacking dedicated experimental evidence of neutral impact emission on our TiN probe, we therefore reason it would be useful to test and evaluate the water vapour impact hypothesis first, extrapolating from what we know from experiments and cometary fly-bys, but allow for uncertainties of up to a factor two, at least.

\begin{figure*}
    \includegraphics[width=1.9\columnwidth]{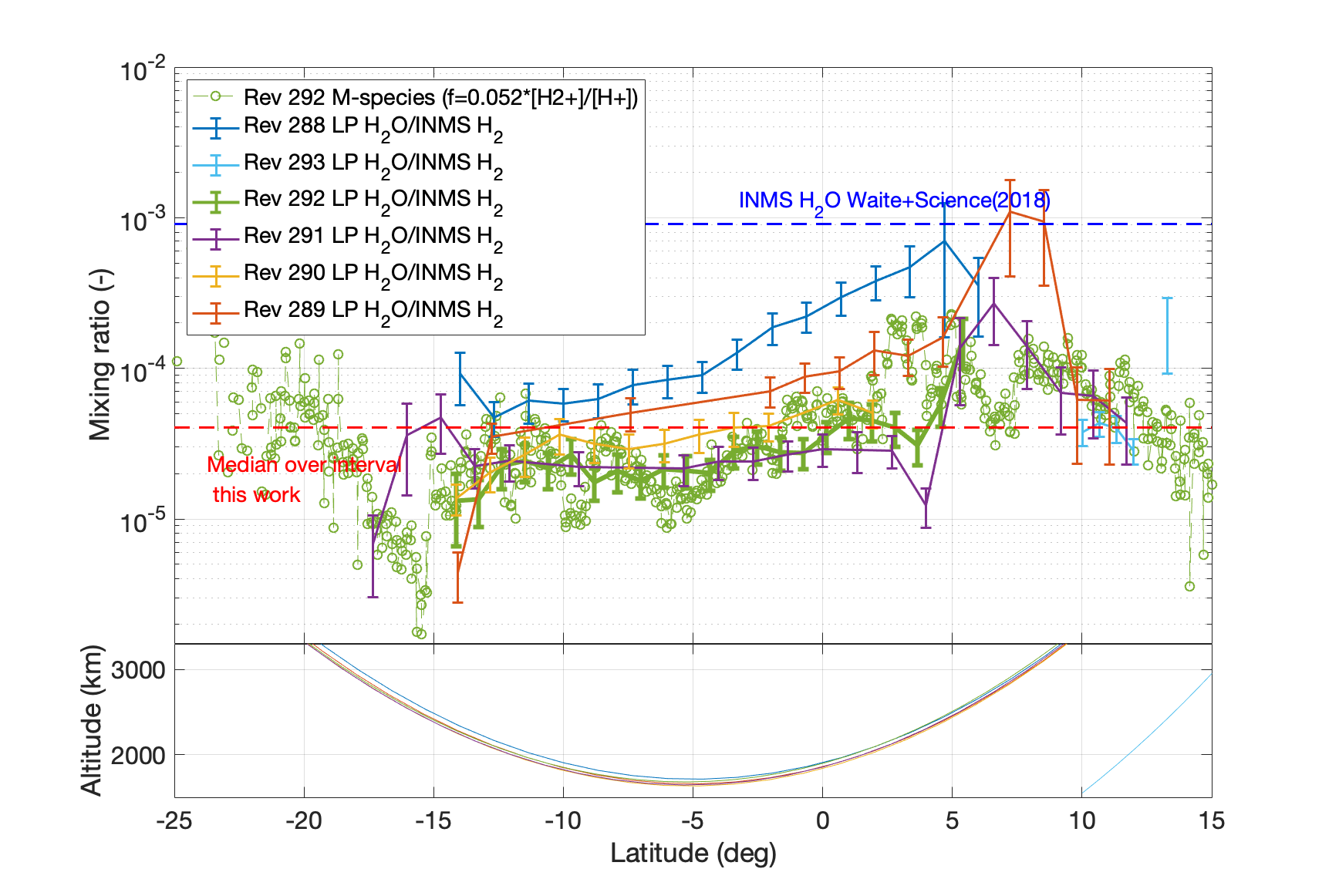}
       \caption{\textbf{Top}:Mixing ratio vs planetocentric latitude, assuming a molecular quantum yield of 0.1 for error bars coloured by revolution for six  different orbits. The error bars are calculated from the error estimate in \ch{H_2} from the INMS as well as a ten percent error in the identification of the secondary electron current. The estimated mixing ratio of an M-type molecule using H2+ and H+ measurements of INMS for Rev. 292 (green circles).  Also plotted as horizontal dashed lines, the median mixing ratio of the six revolutions (288-293) in red, and the mixing ratio reported in \citet{waite_chemical2018science} in blue. \textbf{Bottom}: Planetographic altitude vs planetocentric latitude profile for each orbit, same colouring as above.}

    \label{fig:mixlats}
\end{figure*}

We convert the detected secondary electron emission for each sweep to a neutral density via  Eq.~\ref{eq:ise0} after first removing a 0.5~nA photoemission and a small ion current component assuming $n\ind{i} = n\ind{e}$ from RPWS. As a first approximation, we will use the quantum yield of water on gold from \citet{schmidtarends1985See}, which is the lowest yield (for water gas) of the two metals tested. This can therefore be considered as a generous water gas estimate during the flybys, knowing that contributions from other neutrals will inflate this number. This 'water' density is estimated every 48~s by the RPWS-LP sweep, and we arrive at a mixing ratio by dividing by linearly interpolated \ch{H_2} estimates from INMS and plot the results in Fig.~\ref{fig:mixlats}.

The error bars are dominated by the error estimates in \ch{H_2} from INMS as random errors in the sweep detection algorithm were estimated to be at most 10 percent. However, using \ch{Al} as a proxy for the quantum yield of water on TiN would reduce the water estimate by a factor of two, such that the uncertainty in $Y$ dominates the standard error in our estimate.

In Fig.~\ref{fig:mixlats}, we find that the 'water' mixing ratio varies with up to a factor of 3 between orbits and latitude, and has a distinct evolution during each revolution, showing that at least one of our candidate molecules have a distinct structure in latitude at altitudes between 1500 and 3500 km.

Following \citet{Cravens_ion_comp_2019}, we can set up a simple chemistry model to estimate the mixing ratio of a so-called M-type molecule from the \ch{H^+} and \ch{H_2^+} measurements from INMS. The model relies on the photochemical equilibrium of \ch{H_2^+} and \ch{H^+}, where the loss rate of \ch{H^+} is primarily driven by reactions with M-type molecules (which include \ch{H_2O}, \ch{CH_4} and \ch{CO_2}). Using the methodology described in \citet{Cravens_ion_comp_2019}, we arrive at an estimate for the mixing ratio of M-type molecules, $f_M$, of

\begin{equation}
    f_M = \gamma \frac{k_1}{k_2} \times \frac{[H_2^+]}{[H^+]} \approx 0.052\substack{+0.05 \\ -0.025} \frac{[H_2^+]}{[H^+]},
\end{equation}
where $\gamma$ is the relative production rate of \ch{H_2^+}  to \ch{H^+}, $k_1$, $k_2$ are rate coefficients as defined by \citet{Cravens_ion_comp_2019}, and the errors are estimated from the extreme assumption of only water M-type molecules versus the other extreme, of only \ch{NH_3,~CH_4} and \ch{CO_2} molecules, in combination with uncertainties in gamma and the rate coefficients in literature \citep{Galand2009Solar,UMIST2013,Kim2014saturn}. We note here that the model (1) estimates one mixing ratio for all M-type molecules (which does not detract from a comparison to the LP estimate), (2) depends on a classical quasineutrality assumption ($n\ind{e} = n\ind{i}$), and (3) that the presence of charged dust may change this estimate \citep{Vigren2022Saturn_grain}.

\begin{figure*}
    \includegraphics[width=1.9\columnwidth]{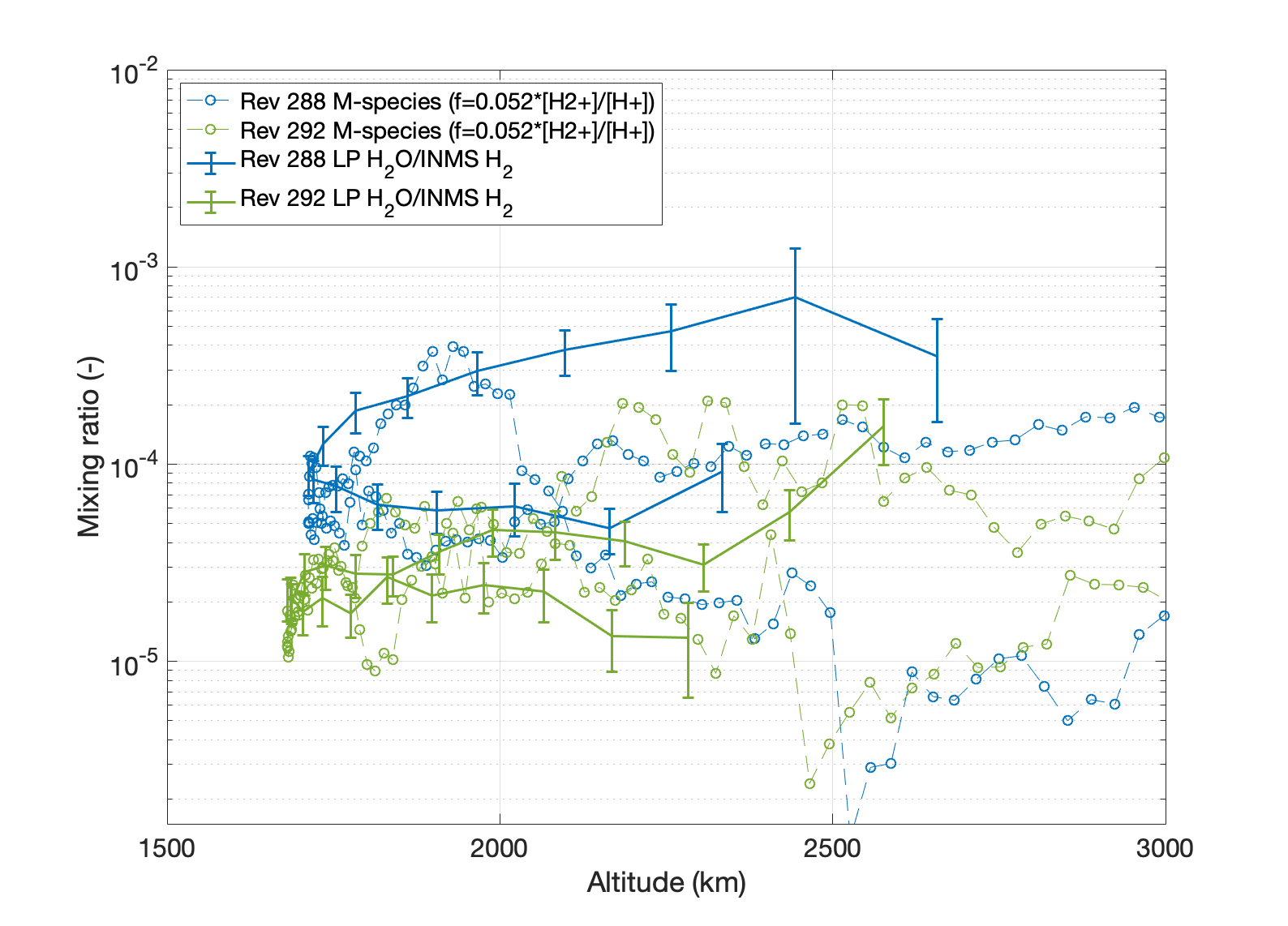}
    \caption{Mixing ratio vs planetographic altitude, assuming a quantum yield of 0.1 for Rev 288 (blue error bars), and Rev 292(green), compared with estimated mixing ratio of an M-type molecule using \ch{H^+} and \ch{H_2^+} measurements of INMS for Rev. 288 (blue circles) and Rev. 292 (green circles). The error bars are calculated from the error estimate in \ch{H_2} from the INMS as well as a ten percent error in the identification of the secondary electron current.}
    \label{fig:h2ochem_alt}
\end{figure*}

We plot the results of that model for Rev 288 and 292, which are the only revolutions where we have INMS \ch{H_2^+} and \ch{H^+} estimates, and compare to our 'water' mixing ratio in Fig.~\ref{fig:h2ochem_alt}. We find a generally very good agreement in magnitude (on a log scale) for large parts around closest approach, both inbound and outbound, but also in the size of the variations. However, Rev. 288's outbound leg shows significant departures from the rest of our estimates, suggesting perhaps that a different molecule (e.g. \ch{N_2}) is responsible for a large fraction of the secondary emission here. We find superficial support for this argument also in \citet{Cravens_ion_comp_2019} (Fig. 4), where 'R-type' molecules such as \ch{N_2} appear to dominate over M-type molecules in a small region past the closest approach of Rev. 288, at the same latitudes where our mixing ratios diverge. 
The general agreement for Rev 292 appears even stronger when plotted versus latitude in Fig.~\ref{fig:mixlats}. From the region where the LP signal is strongest, around -14 to +2 degrees latitude, comparing to the M-species model of the same revolution, we find that the data always agree within a factor of two, but are generally 14 percent below the chemistry model.

\section{Discussion}
\label{sec:discussion}

We start by a detailed discussion of the sweep in Figure~\ref{fig:288_mich}, comparing our analysis including secondary electron emission to the analysis of M2019. This example well illustrates the general differences between the two approaches and the resulting conclusions on the ionospheric environment.

\subsection{LP Sweep Analysis} \label{sec:discussion_sweep}

The secondary electron current, whose amplitude was estimated by the automatic routine, works very well to describe the peak and relaxation of the derivative, as well as the shape of the sweep, assuming $\alpha$=0.12, described by Eq.~\ref{eq:vdagger}. Using the expression for secondaries from Eq.~\ref{eq:ise}, we estimate $T\ind{se}$ to be 0.5~($\pm$0.1)~eV from the relaxation of the derivative after the peak in Fig.~\ref{fig:288_mich} (bottom) around $V\ind{b} = 0$. From the fit of the exponential just below $V\ind{\dagger} = 0$, we arrive at a best estimate of $T\ind{si} =0.1$~eV, and secondary ion yield to be a factor of 10 lower than the electron yield, as anticipated. However, the confidence in the fit of the smaller currents, including $I\ind{si}$ and the ion current $I\ind{i}$ is low owing to the noise level of the sweep.

The spacecraft potential is readily identified in a sweep of a spherical probe as the region where the electron current goes from exponential (electron repulsion from a negative probe) to linear (electron attraction to a positive probe). As the transition point should be where the probe is at the same potential as surrounding space, the bias voltage at that point should be the negative of the spacecraft potential. By this method, it is apparent in Fig.~\ref{fig:288_mich} that $V\ind{s} = 1.6$~V. This differs from the value of 0.16~V suggested by M2019, whose values generally coincide with the peak of the current derivative and the floating potential (i.e.\ the potential at current equilibrium) as is appropriate for a probe in an environment where the electron current dominates at zero potential. In that case, $V\ind{p} = 0$ can be identified as the point where the derivative of the current reaches a maximum, but this is not the case when photoemission or secondary electron emission dominates as this depends on $V\ind{\dagger}$, the potential difference between the probe and its immediate surroundings (as a photoelectron is born at the potential of the probe, but a plasma electron is incident from a plasma at infinity, at a potential of zero per definition).

The slightly positive value of $V\ind{S}$ obtained by M2019 for the sweep in Fig.~\ref{fig:288_mich} is not unreasonable for a spacecraft for the conditions assumed in that paper. Likewise, our roughly ten times as positive spacecraft potential estimate ($V\ind{S}$ = 1.6 $\pm0.2$~V) is reasonable if secondary electron emission dominates the current from the spacecraft at zero potential. Both methods are therefore consistent (or at least not obviously inconsistent) with their underlying assumptions regarding the resulting value of $V\ind{S}$ as well as with the methods used for finding this value.

The effective or harmonic mean mass of ions in Fig.~\ref{fig:288_mich}, assuming the relative velocity is equal to the spacecraft ram speed, is then estimated to be 6.5~amu. The nearest INMS ion density estimate (for ions up to 8~amu) is 650~\cc, with a harmonic mean mass of 1.5~amu. Therefore, our harmonic mean estimate of 6.5~amu suggests that the bulk of the ions not detected by INMS has a mass around 20~amu, and therefore shows more consistency with ionospheric models or in-situ measurements \citep{Cravens_ion_comp_2019} than the analysis of M2019. We note that the distinction between harmonic and arithmetic mean values is here crucial for ion mass estimates from LP data.

The electron current, estimated from a linear least-square fit above $V\ind{b} > 1$~V, and a small exponential contribution detected below $V\ind{p}$=0 ($V\ind{b}<$-1.6 V), allows us to model a linear current contribution that was proposed in M2019 to be from a large population of negative ions with a speed very close to the spacecraft velocity as instead due to the electrons. Technically, the LP cannot distinguish an electron from a heavier, but very cold and slow negative particle. However, as there is no possibility to confirm the presence of these negative ions by other instruments and a model for creating large amount of negative ions with these low relative velocities was not presented, our interpretation in terms of ionospheric electrons is more conservative. We may also note that in order to model this current, M2019 used a spacecraft potential of about $\approx$ 2~V for this negative ion current, differing from the value of 0.16~V used for all other currents. 

However, our approach is not without complications. For Fig~\ref{fig:288_mich} (but also generally), we find a rather high $T\ind{e}$ (0.45 $\pm$ 0.15~eV) and a lower electron density than M2019 (2410~\cc), much lower than the nearest RPWS wave measurement of 3740~\cc. In principle, this leaves room for a significant charged dust population to ensure quasineutrality, as concluded in M2019. However, this interpretation has issues of its own, which we will return to in the next Section~\ref{sec:dustsensitivity}.

One should remember that the LP estimates, which are sampled from a small volume immediately surrounding a probe close to the spacecraft, may be different from the plasma parameter in an undisturbed ionosphere. In contrast to fast-flowing ions, neutrals and dust, plasma electrons could readily be perturbed from their orbit to the LP instrument by spacecraft-plasma interactions \citep{odelstad_measurements_2017,Johansson2020charging,Johansson_plasmaxcal_2021,zhang}. As the spacecraft is charged (in V) much above the electron energy (in eV), we cannot rule out that this could significantly affect our measurements. In particular, the cold (0.04~eV) ionospheric electrons we expect in the deep ionosphere \citep{Moore2008temperaturesSaturn} would be perturbed even further, or may not be able to reach our probe, for instance if an electron density enhancement in front of the spacecraft has formed \citep{ivchenko_disturbance_2001}.

That the RPWS-LP sweep electron densities are often significantly lower than the RPWS data is not new, and is also evidenced by comparing the nearest RPWS estimates with the published LP sweep analysis in M2019. However, since the sweep and the 20~Hz current data are internally consistent (i.e. the currents at the potential measured in 20~Hz resolution matches the nearest sweep current measured at that potential), the ad-hoc scaling of the 20~Hz data to electron densities that were validated by calibration to the RPWS electron density in M2019, will of course work also for the RPWS-LP sweep electron density estimates. That a scaling mitigates the issue is also consistent with our hypothesis that electrons below a certain energy cannot reach the probe, as the fraction of the density that reaches the probe will then be constant even if the number density fluctuates for a Maxwell-Boltzmann energy distribution of electrons. Moreover, the LP estimated electron temperature would then be inflated, as it would only reflect the energies of the electrons that reach our probe.

The total fits presented in Figs.~\ref{fig:288_mich}, \ref{fig:291}, \ref{fig:292_14284} \& \ref{fig:288_id386} have a few departures from the measurements. Although the high noise level complicates the evaluation, we are aware of potential deviations from our model assumptions of secondary emission currents (Eq.~\ref{eq:ise} \& .~\ref{eq:isi}). We suspect that the largest errors arise from our assumption of a Maxwellian energy distribution of the emitted particles, as discussed in Sec.~\ref{sec:method}, which is the origin of the sharp feature in the derivative (e.g. at $V\ind{b}=$-0.16~V in Fig.~\ref{fig:288_mich}). There might be better models for emission (such as a logistic regression), but for this study, even errors of order 10 percent will not significantly affect our interpretation. We hope to restrict the number of free parameters by not adding more current contributions than we deem strictly necessary to reduce the residual below 5 percent. All in all, we find that our fit in Fig.~\ref{fig:288_mich} reduces the residual with two orders of magnitude compared to M2019.

In contrast to \citet{Garnier2013SEE}, we do not see the tell-tale negative slope in the sweep to suspect secondary electron emission from \emph{electron} impact, indicating instead a neutral source particle. The electron spectrometer CAPS-ELS was not operational during the final revolutions to estimate the high energetic electrons directly, but the electron differential energy flux needed to produce the secondary current we measure would produce a rather impressive aurora at $>10^{10}~($keVsrcm$^2$s$)^{-1}$, within a few degrees of the equator for all fly-bys. In absence of such auroral observations, and since \citet{Garnier2013SEE}(Fig 2.) predicts this current not to be relevant for the LP at these magnetic latitudes, we ignore this effect in our analysis.

\subsection{LP Sensitivity to charged dust}
\label{sec:dustsensitivity}
As the impact energy of dust is very large, (above several keV for even the smallest grains), the current from charged dust can be assumed to be constant over the sweep (a few V), and estimated using the formalism of Eq.~\ref{eq:Ii0} for a continuous stream of dust particles. For the sweeps presented in Section ~\ref{sec:results} and in the Appendix~\ref{sec:appendix}, no residual offset and thus no detectable dust component could be identified. Nominally, this suggests that the charged dust density will need to be below 10~cm$^{-3}$ to avoid being detected, using the noise level of 0.1~nA reported in~\citep{gurnett_cassini_2004}, although with the noise level present in these sweeps, a more realistic lower limit is 100~cm$^{-3}$.

Nevertheless, second order effects of dust impinging the probe would play a significant role. As the dust impact velocity is much larger than the sound speed of any spacecraft material, we move into the realm of high-velocity dust detection. Here the impactor and spacecraft material vaporises and ionises, forming an expanding plasma cloud and a subsequent large spike in current \citep{Schippers_nanodust2014,MeyerVernet_nanodust_stereo2009}.

This would be detectable also in the RPWS data but has not been reported at these altitudes. This argument, based on kinetic energy, is also central in another point of contention with the published analysis of M2019, inferring a large (up to 50.000~\cc) charged dust population but lacking direct detection of them. As a local electric potential needed to screen the Langmuir Probe from charged dust would need to be on the same order of magnitude in Volts as their impact energy in eV, i.e. thousands or millions of Volts, even the lightest dust grain will be harder to deflect than any observed ionospheric electron.

There are sweeps that are particularly noisy, perhaps signatures of non-continuous dust impact spikes, much like the sweeps near the Enceladus plumes. However, these instances are rare in the ionosphere of Saturn, and would warrant a more focused study.

\subsection{Neutrals}
\label{sec:discussion_neutrals}

Despite the many assumptions, our first approximation of water vapour impact emission yields an excellent agreement with our chemistry model in Fig.~\ref{fig:mixlats}.

Choosing water on gold as a proxy for the secondary electron emission for the Cassini LP is seemingly not only consistent with the variation of the signal we see along the orbit as predicted by the variations of H+/H2+ detected by INMS, but is also a better descriptor of relatively large  $I_{sei}/I_{see}$ ratio seen in Figs.~\ref{fig:288_mich},\ref{fig:291},\ref{fig:292_14284} \& \ref{fig:288_id386}, than predicted by e.g. water on aluminium in \citet{schmidtarends1985See}.

The neutral gas profile yield some variations between orbits and altitudes, but with a surprisingly clear and commmon structure in latitude. In absolute numbers, we note that the concentrations from the deep ionosphere given in \citet{waite_chemical2018science}, yields a water gas mixing ratio of $9\times10^{-4}$, but only three LP estimates reach these values. These specific estimates appear as outliers, as they are a factor of 30 above our median estimate.
Since $Y^e\ind{H2O}$ is relatively well known, we do not find it reasonable to stretch an already generous LP water gas estimate (assuming no secondary emission from \ch{CH_4,~CO_2,~N_2}, etc) with a factor of 30.

The chemistry model is independent of LP measurements (and errors) and we could reasonably argue for errors of a factor of two, although an extension of the same model predicts the RPWS electron densities with a good match \citep{Cravens_ion_comp_2019}. Even so, a factor of two is not enough to support such a water mixing ratio of $9\times10^{-4}$ for any M-type molecule, and is as such not compatible with the water content from INMS \citep{waite_chemical2018science}.

Instead, we suggest that INMS detects not only the water that enters the instrument as water vapour, but also the water created when icy particles impact inside the instrument and vaporise. In fact, \citet{miller_cassini2020} (Fig. 4) compares the volatiles in the final four orbits, including the final plunge, and find that the final plunge is depleted in \ch{H_2O}, but also \ch{NH_3} and \ch{CO_2}. This was seen as a clear indication that water signal in INMS is carried by watery/icy grains that are seemingly predominant in the equatorial plane \citep{miller_cassini2020,Perry2018Saturnupper} which is not sampled in the final plunge, and seemingly very consistent with this work. The possibly inflated atmospheric water vapour estimate in INMS would also be amplified at these low altitudes as the differential rotation of the grains due to collisions with the \ch{H_2} atmosphere becomes significant and prompts watery grains to vaporise more rapidly inside the instrument chamber.

We note that the chemistry model is only applicable for water vapour and thus not sensitive to such grains. Similarly, the LP impact emission is only tested for molecules, and non-continuous/quasi-continuous impacts of small water-rich grains may appear as spikes or noise in the sweep, which is ignored in the analysis, but may carry enough water content to explain the discrepancy between INMS and our estimates.

\section{Conclusions}\label{sec:conclusion}

The secondary emission as detected by RPWS-LP is consistent with an abundant and varying neutral molecules abundance, of which detection of water vapour appears most likely, but can also be significantly modified by variations of other "M-type" molecules such as \ch{CH_4} or \ch{CO_2}. The mixing ratio varies smoothly, with generally consistent latitude trends for all fly-bys. These variations exceed one order of magnitude between 2500 and 1500~km altitude for all orbits. There are also an order of magnitude variations between orbits and latitude, with the highest mixing ratio detected during revolution 288 and 289, indicating a strongly varying chemical composition in the Saturnian ionosphere. 

The LP was not specifically designed to measure neutrals in the Saturnian environment. Therefore, we have no specific calibration to rely upon and cannot exclude instrumentational errors. It is clear that the dense atmosphere gives rise to a strong secondary electron current on the Langmuir Probe, although the concentrations of individual molecular species is at the moment unclear. Even so, a simple chemistry model suggests that the dominant origin is most likely an "M-type" molecule, for which our best estimate for the quantum yield of this/these molecules is 0.08 $\pm$ 0.04, which is consistent with lab measurements of water incident on gold, as well as \ch{CO_2}. Future laboratory experiments of \ch{H_2O,~CH_4,~CO_2,~N_2} on space-weathered TiN would allow us to provide a more accurate estimate of each of these neutrals in the Saturnian ionosphere. Also, if this analysis is extended to the Cassini fly-bys of the plumes of Encaladus and the dense ionosphere of Titan, it would allow us to estimate the impact emission yield of a vastly different neutral environment, which in turn helps us constrain the Saturnian ionosphere composition. At Titan and Enceladus, the low fly-by speeds ($\sim$ 5 - 18\kms) will likely render lighter neutral species (including \ch{H_2O}) invisible for the LP. However, the reported atmosphere composition still provides a significant population of molecules >40~Da at Titan \citep{Cui2009Titan}, and of \ch{CO_2} at Enceladus \citep{Teolis2017Astb_enc} which may trigger significant secondary emission. Although we expect the effect on the sweep analysis to be much more modest at these fly-bys (with little to no effect on the 20~Hz LP densities), there is still room for misidentification of electron temperatures, spacecraft potential and overestimation of electron and ion densities.

Here we build and improve upon the sweep analysis of M2019. In particular, the ion density estimate is significantly reduced and lends support for a quasineutral picture of the ionosphere from a balance of electrons and ions. We cannot rule out the presence of charged dust, but for the sweeps presented here, the RPWS-LP measurements are consistent with charged dust densities below 100~\cc. To confirm a significant abundance of charged dust in the ionosphere of Saturn, we would welcome a thorough reanalysis of the RPWS-LP sweeps, taking into account both the current from impact emission as well as the current these charged dust grains would provide to the probe.

$T\ind{se}$ was found to be varying between 0.1-0.6 eV for the environment probed by the RPWS-LP in the Saturnian ionosphere. However, irregularities in the sweep derivative, albeit noisy, suggested that the energy distribution of emitted electrons is not perfectly Maxwellian. We therefore welcome further research into how to model such currents, as well as a more detailed laboratory study in the $T\ind{se}$ variation with chemical composition of the neutral gas and impact energy.

This and future studies into impact emission is especially important for the upcoming Comet Interceptor mission, where a similar instrument would be able to characterise the cometary neutral environment via impact emission if the fly-by velocity is large enough (the target is not selected before launch, in 2029).

\section*{Acknowledgements}
This work would not have been possible without the collective efforts over a quarter of a century of all involved in the \emph{Cassini} project and RPWI. FLJ acknowledges the support from the Lennanders Stipend. JD is thankful for being supported by the Swedish National Space Agency under grant Dnr 143/18. JHW and KM received funding from JPL's Cassini Project Office during the collection of the data and from internal funding from Southwest Research Institute for the post mission analysis described herein.

\section*{Data availability}
The RPWS-LP Sweep data underlying this article are available in PDS, at \url{https://doi.org/10.17189/1519611}. All other data underlying this article will be shared on reasonable request to the corresponding author.

\bibliographystyle{mnras}
\bibliography{References_all.bib}

\appendix

\section{Sheath-Limited Theory}\label{sec:appendix}
When $r_p/\lambda_D << 1$ no longer holds, the OML assumptions for attracting current are no longer valid ( although it is generally a very good approximation even for $r_p = \lambda_D$), and we will need to consider Sheath-Limited Theory (SL). \citet{laframboise_theory_1966} solved  the particle orbits and currents to a probe for such a plasma numerically. In a plasma where the attracted species (here: electrons) are more mobile (and therefore sets the debye length), the electron current to an attracting spherical probe can be calculated from \citep{laframboise_theory_1966} Table 6c (also represented as a graph in Figure 20), depending on $r_p/\lambda_D$ where $r_p/\lambda_D = 0$ is the OML solution, and serves as an upper estimate of the current at any other debye lengths. For Fig~\ref{fig:288_mich}, using M2019 estimate of the spacecraft potential and $r_p/\lambda_D $ = 0.3, assuming $T\ind{e} $ = 0.4~eV, $n\ind{e}$ = 3740~\cc, the expected overestimation from OML is much less than a percent at $V\ind{b}$=3~V. However the (OML) M2019 analysis diverges from the measurements by more than 100 percent at $V\ind{b}$=3~V. Even the extreme estimate of $T\ind{e}$ = 0.05~eV cannot explain more than an error of 10 percent at $V\ind{b}$=+1~V, where the actual discrepancy is three times larger.

\begin{figure}
    \includegraphics[width=1.0\columnwidth]{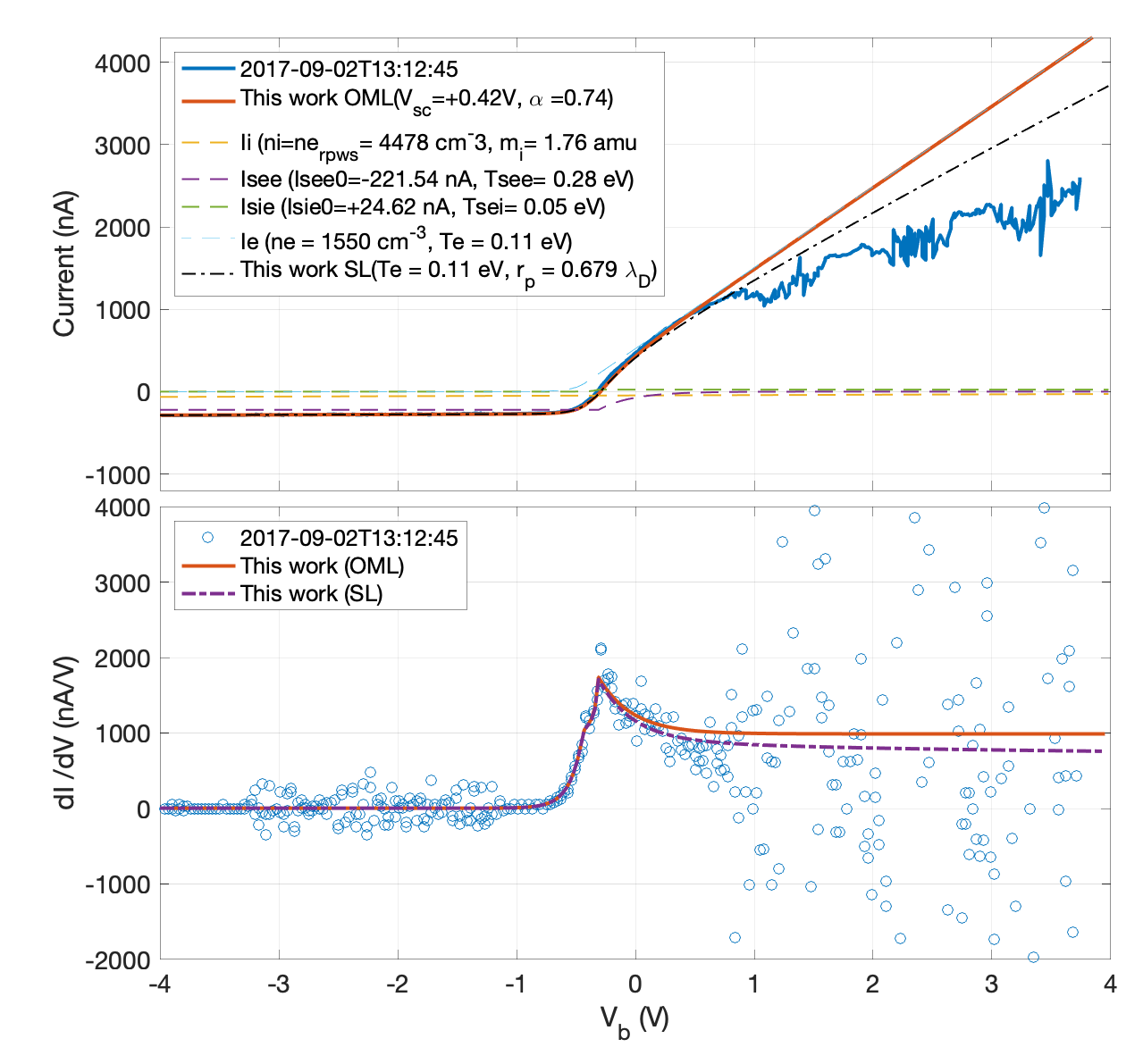}
    \caption{Currents and current derivative vs bias potential, common x-axis.\textbf{Top:} Langmuir Probe sweep from 2017-09-02T13:12:45 from revolution 291 (blue line) and fitted sweep currents, including secondary electron and ion emission from neutrals (purple and green dashed line, respectively), electrons (blue dashed line), ions (yellow dashed line) and the total fit according to Orbital-Motion Limited Theory (red solid line) and Sheath Limited Theory (black dot dashed  line) vs bias potential.  \textbf{Bottom:} The current derivative of the sweep (blue circles), of the total fit of the analysis according to Orbital-Motion Limited Theory (red solid line) and Sheath Limited Theory (purple dot dashed line) vs bias potential.}
    \label{fig:291}
\end{figure}

\begin{figure}
    \includegraphics[width=1.0\columnwidth]{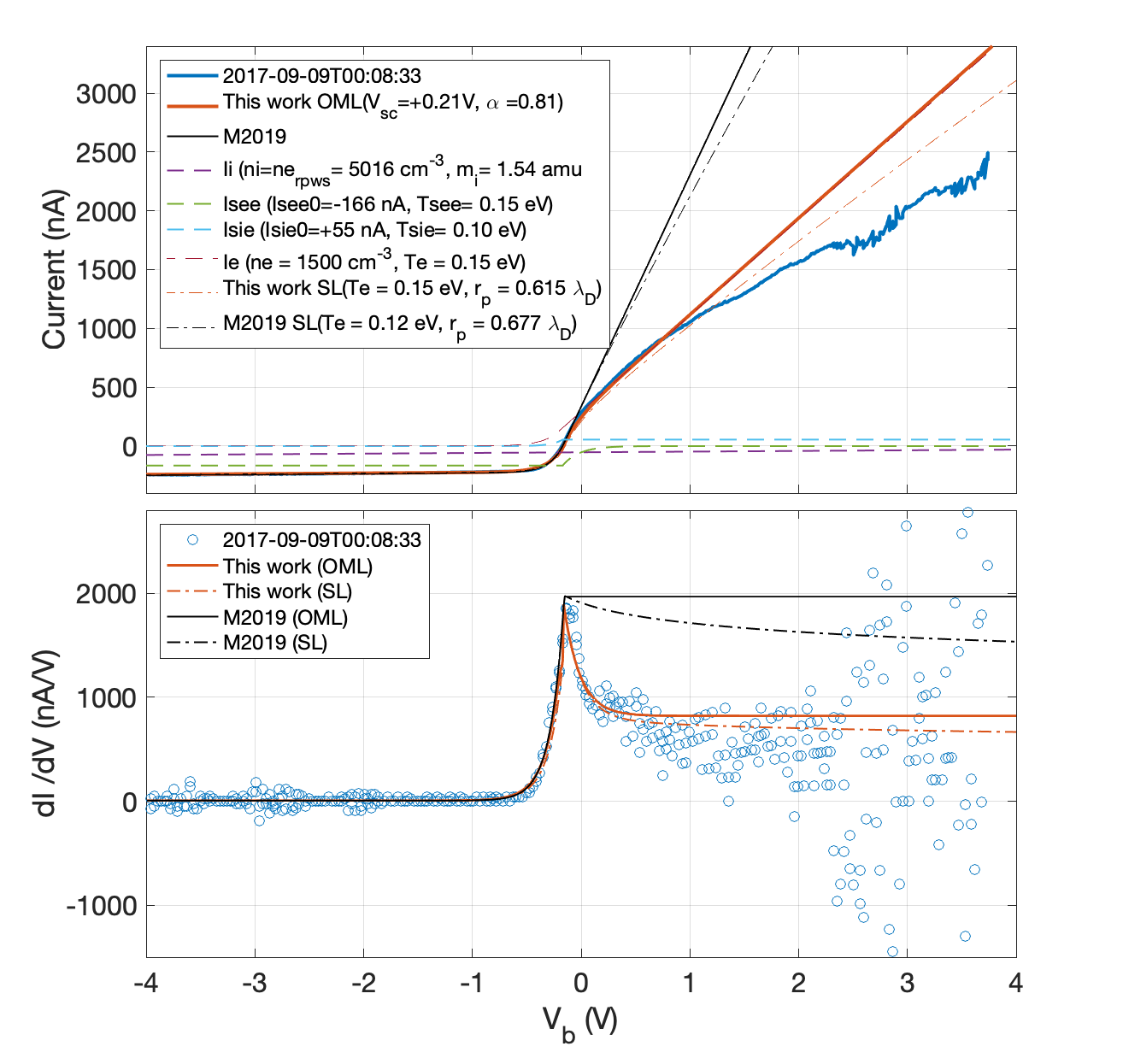}
    \caption{Currents and current derivative vs bias potential, common x-axis.\textbf{Top:} Langmuir Probe sweep from 2017-09-09T00:08:33 from revolution 292 (blue line) and fitted sweep currents, including secondary electron and ion emission from neutrals (green and light blue dashed line, respectively), electrons (red dashed line), ions (purple dashed line) and the total fit according to Orbital-Motion Limited Theory (red solid line) and Sheath Limited Theory (red dot dashed line) vs bias potential. Also shown, the fit of the same sweep published in M2019, and the corresponding Sheath-limited adaption of the same sweep (solid and dot-dashed black line, respectively). \textbf{Bottom:} The current derivative of the sweep (blue circles), of the total fit of the analysis in this work (in red) and in M2019 (in black) according to Orbital-Motion Limited Theory (solid line) and Sheath Limited Theory (dot dashed line) vs bias potential.}
    \label{fig:292_14284}
\end{figure}

\begin{figure}
    \includegraphics[width=1.0\columnwidth]{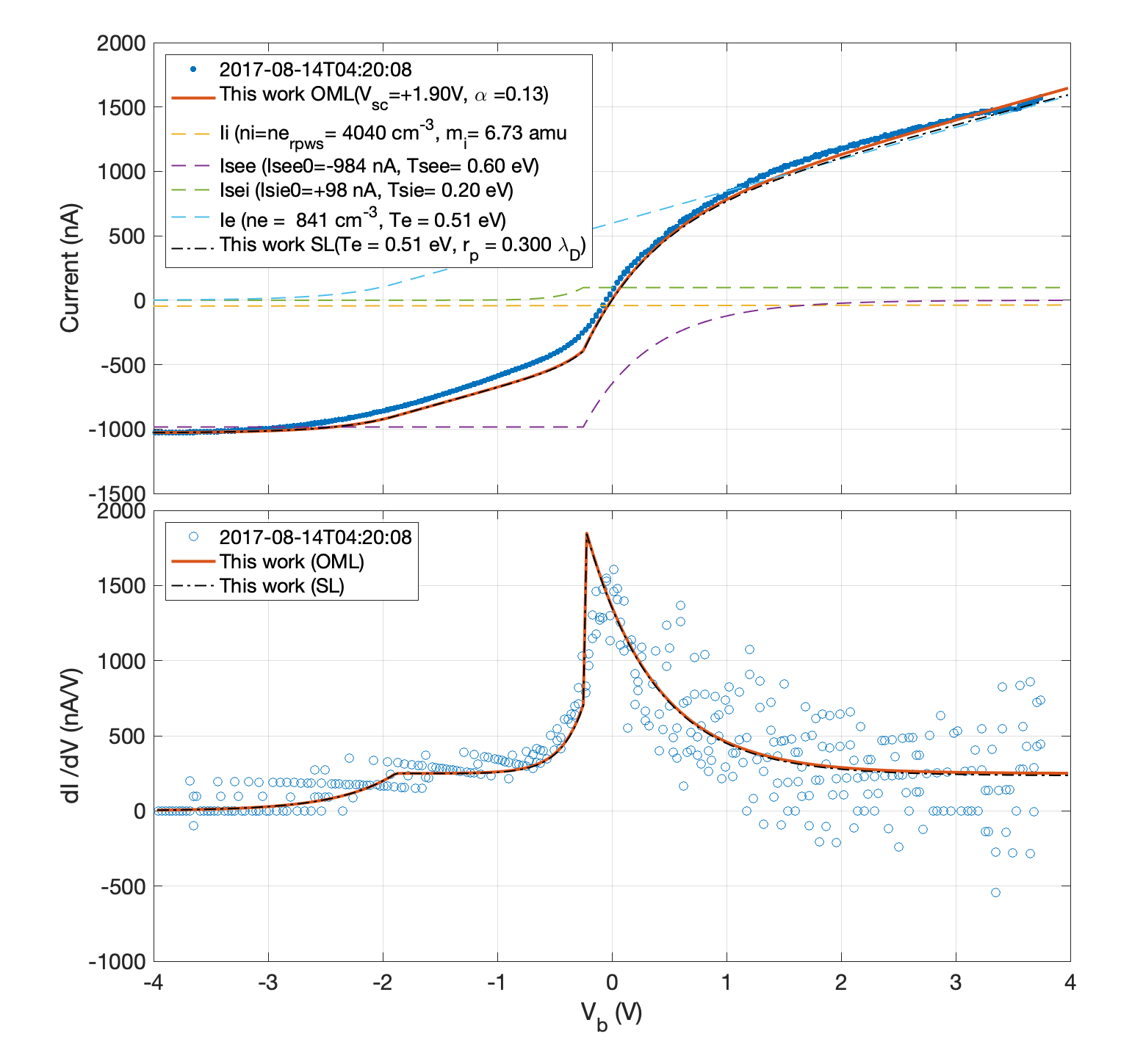}
    \caption{Currents and current derivatives vs bias potential, common x-axis.\textbf{Top:} Langmuir Probe sweep from 2017-08-14T04:20:08 from Rev. 288 (blue dots) and fitted sweep currents, including secondary electron and ion emission from neutrals (purple and green dashed line, respectively), electrons (blue dashed line), ions (yellow dashed line) and the total fit in OML (red solid line) and SL( black dot-dashed line) vs bias potential. The harmonic mean of the ion mass from INMS (measuring ions up to 8~amu) at the time of the sweep was 1.73~amu. \textbf{Bottom:} The current derivative of the sweep (blue circles), of the total fit of the analysis in OML (red solid line) and in SL (black dot-dashed line) vs bias potential.}

    \label{fig:288_id386}
\end{figure}

Interpolating the tabulated values in \citet{laframboise_theory_1966} (Table 6c), we estimated the SL current and derivative of the sweep also in the densest regions, as seen in Fig.~\ref{fig:291} for Rev. 291 at an altitude of 1673~km, in Fig.~\ref{fig:292_14284} for Rev. 292 at an altitude of 1718~km, and in Fig.~\ref{fig:288_id386} for Rev 288, at an altitude of 1740~km, all near closest approach. We note first of all that in the absence of instrumental noise, wave activity, spacecraft potential or plasma density fluctuations on the timescale of the sweep (0.5~s), we cannot explain a negative current derivative as it would imply a negative resistance in the plasma. However, a negative derivative is indeed seen at  $V\ind{b} = 0.8$~V for Fig.~\ref{fig:291} and at $1.2~$V for Fig.~\ref{fig:292_14284}. It is therefore likely that we are seeing also a temporal or, since the spacecraft is moving, spatial effect. For instance, the large current to the probe (which is electronically connected to the spacecraft circuitry) can have shifted the spacecraft potential during the sweep. Given the somewhat sinusoidal (albeit erratic) shape beyond 1~V, we can also not rule out electron density fluctuations from a plasma wave carried by electrons, or other density fluctuations. A mitigating strategy would then be to limit our analysis to $V\ind{b}<0.8~$V for these sweeps, as the total amplitude of current is smaller, and the wave amplitude would also be negligible when less plasma electrons contribute to the current. The accuracy of the fit below $V\ind{b}<0.8~$V speaks in favour of this strategy, and since one of these disturbed sweeps was included in M2019 (Fig 3), we feel it is still useful to discuss and compare.

The application of Sheath-Limited theory does slightly improve the fit at larger potentials, as expected, but cannot explain the relaxation in the derivative alone. This is apparent in all sweeps in the deep ionosphere, but perhaps most striking when applied to the M2019 analysis without secondary electron emission as shown by the black dot-dashed line in Fig.~\ref{fig:292_14284} (Top and bottom).

The sweep analysis of Fig.~\ref{fig:291}, ~\ref{fig:292_14284} \& \ref{fig:288_id386} all suggests that the LP estimates may not reflect the undisturbed plasma in Saturn's ionosphere as the electron density is lower than the RPWS estimate, and the electron temperature is warmer than what we expected \citep{Moore2008temperaturesSaturn}. Moreover, we cannot confirm a departure from classic quasi-neutrality (i.e. $n\ind{i} = n\ind{e}$ from RPWS) and we fail to detect a current from charged dust grains.

\bsp	
\label{lastpage}
\end{document}